\documentclass[a4paper,12pt]{article}
\usepackage[english]{babel}
\usepackage{graphicx,color}
\usepackage{amsmath,amsthm,amssymb,amsfonts}
\usepackage{epsfig}
\usepackage[T1]{fontenc}
\usepackage[cp1250]{inputenc}
\usepackage[bookmarks,colorlinks]{hyperref}
\usepackage{fancyhdr}
\fancypagestyle{plain}{%
\fancyhead{}

}

\hypersetup{colorlinks,%
linkcolor=magenta,%
urlcolor=blue,%
citecolor=red}

\newcommand{\tr}{\mathrm{Tr}}
\newcommand{\sgn}{\mathrm{sgn}}
\newcommand{\artanh}{\mathrm{artanh}}
\newcommand{\cor}{\mathrm{Cor}}
\newcommand{\vac}{\left|\mathrm{VAC}\right\rangle}
\newcommand{\ivac}{\left\langle\mathrm{VAC}\right|}

\title{Global One--Dimensionality conjecture within Quantum General Relativity}

\author{
\textbf{\L ukasz Andrzej Glinka}\footnote{Previously at Bogoliubov Laboratory of Theoretical Physics of Joint Institute for Nuclear Research in Dubna, Russian Federation}\vspace*{15pt}\\
E-mail: \href{mailto:laglinka@gmail.com}{\bf{\tt{laglinka@gmail.com}}}
\vspace*{10pt}\\
\emph{International Institute for Applicable}\\
\emph{Mathematics \& Information Sciences,}\\
\emph{Hyderabad (India) \& Udine (Italy),}\vspace*{10pt}\\
\emph{B.M. Birla Science Centre,}\\
\emph{Adarsh Nagar, 500 063 Hyderabad, India}
}
\date{\today}

\begin{document}

\maketitle
\thispagestyle{empty}
\begin{abstract}
The simple quantum gravity model, based on a new conjecture within the canonically quantized $3+1$ general relativity, is presented. The conjecture states that matter fields are functionals of an embedding volume form only, and reduces the quantum geometrodynamics. By dimensional reduction the resulting theory is presented in the form of the Dirac equation, and application of the Fock quantization with the diagonalization procedure yields construction of the appropriate quantum field theory. The 1D wave function is derived, the corresponding 3-dimensional manifolds are discussed, and physical scales are associated with quantum correlations.\\

\noindent \textbf{Keywords} general relativity; 3+1 splitting ; quantum gravity models ; low dimensional quantum field theories ; quantization methods ; global one-dimensionality\\

\noindent \textbf{PACS} 04.60.-m ; 05.30.Jp ; 05.70.Ce; 11.10.Kk ; 98.80.Qc
\end{abstract}
\newpage
\section{Introduction}

Quantum gravity is one of the fundamental problems of modern theoretical physics. In spite of the significant efforts and various approaches, we are still very far of understanding the role of quantized gravitational fields in physical phenomena at high energies (for different approaches to quantum gravity see \emph{e.g.} Ref. \cite{qg}). In this paper we propose a simple model of quantum gravity which can be useful for clarifying its some important aspects.

The celebrated field-theoretic formalism yields plausible phenomenology for numerous experimental data of all areas of physics. In this paper this point of view is used for construction of a simple quantum gravity model. The $3+1$ splitting of a general relativistic metric tensor and the canonical quantization of the appropriate action functional are used in the well-grounded way. In straightforward and strict analogy with the generic cosmological model \cite{gli}, the new conjecture within the Wheeler--DeWitt quantum geometrodynamics is proposed. The model is based on the ansatz composed by the steps
\begin{enumerate}
  \item global one--dimensionality conjecture, \emph{i.e.} one-dimensional matter fields,
  \item reduced quantum geometrodynamics, yielding one-dimensional theory,
  \item dimensional reduction, resulting in the Dirac equation formulation,
\end{enumerate}
and expressing the supposition that the quantum geometrodynamics in itself is a one-dimensional field theory. The dimensional reduction leads to the appropriate Dirac equation and the Euclidean Clifford algebra. Its the Fock quantization with the diagonalization procedure, consisting of the Bogoliubov transformation and the Heisenberg equations of motion, yields correctly defined quantum field theory. The resulting model describes quantum gravity in terms of a quantum field theory formulated in the Fock static operator reper associated with initial data. The 1D wave function is derived, the corresponding 3-dimensional manifolds are discussed, and quantum correlations are associated with physical scales. Mathematically, we employ the one-dimensional functional integrals, so that the proposing quantum gravity model is methodologically corresponding to the trend initiated by Hartle and Hawking in the paper \cite{haw}.

An organization of the paper is as follows. In the preliminary section \ref{sec:1} historically first quantized 3+1 general relativity is presented. Section \ref{sec:2} is devoted to the ansatz presentation. Next, the sections \ref{sec:3} and \ref{sec:4} discuss field quantization and some implications of general formulation, respectively. Finally, in the section \ref{sec:5} the entire paper's results are summarized in condensed way.
\section{Quantum 3+1 General Relativity}\label{sec:1}
In general relativity (See \emph{e.g.} \cite{gr}) a pseudo-Riemannian manifold $(M,g)$ with a metric tensor $g_{\mu\nu}$, the Christoffel symbols $\Gamma^\rho_{\mu\nu}$, the Riemann curvature $R^\lambda_{\mu\alpha\nu}$, the Einstein curvature $G_{\mu\nu}=R_{\mu\nu}-\frac{1}{2}g_{\mu\nu}{^{(4)}}\!R$, where $R_{\mu\nu}=R^\lambda_{\mu\lambda\nu}$, ${^{(4)}}\!R=R^\kappa_\kappa$, satisfying the Einstein field equations\footnote{We use the units $c=\hbar=8\pi G/3=1$ in this text.}
\begin{equation}\label{feq}
G_{\mu\nu}+\Lambda g_{\mu\nu}=3T_{\mu\nu},
\end{equation}
where $\Lambda$ is cosmological constant and $T_{\mu\nu}$ is stress-energy tensor, models a spacetime\footnote{In (\ref{feq}) the coefficient of $T_{\mu\nu}$ usually equals $8\pi G/c^4$ that is exactly 3 in the our units.}. For a compact $M$ with a boundary $(\partial M, h)$ and curvature $K_{ij}$, the usual variational principle is corrected \cite{ygh} and (\ref{feq}) arise by the action
\begin{equation}\label{eh0}
S[g]=\int_{M}d^4x\sqrt{-g}\left\{-\dfrac{1}{6}R+\dfrac{\Lambda}{3}\right\}+S_\psi[g]-\dfrac{1}{3}\int_{\partial M}d^3x\sqrt{h}K,
\end{equation}
where $K=h^{ij}K_{ij}$, $S_\psi[g]$ is Matter fields action, $T_{\mu\nu}=-\frac{2}{\sqrt{-g}}\frac{\delta S_\psi[g]}{\delta g^{\mu\nu}}$. For $\Lambda=0$, a global timelike Killing field on $M$ $\mathcal{K}$ exists, the foliation $t=const$ is spacelike, $\partial M$ is the Nash embedding \cite{nash}, and $3+1$ splitting \cite{3+1} holds
\begin{eqnarray}\label{dec}
g_{\mu\nu}=\left[\begin{array}{cc}-N^2+N_iN^i&N_j\\N_i&h_{ij}\end{array}\right]\quad,\quad h_{ik}h^{kj}=\delta_i^j\quad,\quad N^j=h^{ij}N_i
\end{eqnarray}
For $\Lambda>0$, $\mathcal{K}$ does not exist, spacelike $\partial M$ only foliates an exterior to the horizons on geodesic lines, (\ref{dec}) is a gauge. In both cases (\ref{eh0}) takes the form
\begin{eqnarray}\label{gd}
  S[g]=\int dt\int_{\partial M} d^3x\left\{\pi\dot{N}+\pi^i\dot{N_i}+\pi_\psi\dot{\psi}+\pi^{ij}\dot{h}_{ij}-NH-N_iH^i\right\},
\end{eqnarray}
where dot means $t$-differentiation, $H$ and $H^i$ are defined as
\begin{eqnarray}
H=\sqrt{h}\left\{K^2-K_{ij}K^{ij}+ {^{(3)}R}-2\Lambda-6\varrho\right\}, \qquad H^i=-2\pi^{ij}_{~;j}~,\label{con2}
\end{eqnarray}
with ${^{(3)}R}=h^{ij}R_{ij}$, $\varrho=n^{\mu}n^{\nu}T_{\mu\nu}$, $n^\mu=[1/N,-N^i/N]$, and particularly
\begin{equation}\label{mom}
\pi^{ij}=-\sqrt{h}\left(K^{ij}-h^{ij}K\right).
\end{equation}
The curvature $K_{ij}$ satisfies the Gauss--Codazzi equations \cite{gd}
\begin{equation}
  2NK_{ij}=N_{i|j}+N_{j|i}-\dot{h}_{ij}.\label{con0}
\end{equation}
where stroke means intrinsic covariant differentiation. Time-preservation \cite{dir} of the primary constraints \cite{dew} leads to the secondary ones -- (scalar) Hamiltonian constraint yielding dynamics, and (vector) diffeomorphism one merely reflecting spatial diffeoinvariance
\begin{equation}
\pi\approx0\quad,\quad\pi^i\approx0 \quad\longrightarrow\quad H\approx0\quad,\quad H^i\approx0.\label{const}
\end{equation}
DeWitt \cite{dew} showed that $H^i$ generate the diffeomorphisms $\widetilde{x}^i=x^i+\xi^i$
\begin{eqnarray}
i\left[h_{ij},\int_{\partial M}H_{a}\xi^a d^3x\right]&=&-h_{ij,k}\xi^k-h_{kj}\xi^{k}_{~,i}-h_{ik}\xi^{k}_{~,j}~~,\\
i\left[\pi_{ij},\int_{\partial M}H_{a}\xi^a d^3x\right]&=&-\left(\pi_{ij}\xi^k\right)_{,k}+\pi_{kj}\xi^{i}_{~,k}+\pi_{ik}\xi^{j}_{~,k}~~,
\end{eqnarray}
and consequently the first-class constraints algebra can be derived
\begin{eqnarray}
  i\left[H_i(x),H_j(y)\right]&=&\int_{\partial M}H_{a}c^a_{ij}d^3z,\label{com1}\\
  i\left[H(x),H_i(y)\right]&=&H\delta^{(3)}_{,i}(x,y),\label{com2}\\
  i\left[\int_{\partial M}H\xi_1d^3x,\int_{\partial M}H\xi_2d^3x\right]&=&\int_{\partial M}H^a\left(\xi_{1,a}\xi_2-\xi_1\xi_{2,a}\right)d^3x.\label{com3}
\end{eqnarray}
Here $H_i=h_{ij}H^j$, and $c^a_{ij}=\delta^a_i\delta^b_j\delta^{(3)}_{,b}(x,z)\delta^{(3)}(y,z)-(x\rightarrow y)$ are structure constants of diffeomorphism group. The scalar constraint reduced by (\ref{mom}) with using of the canonical primary quantization \cite{dir,fad}
\begin{eqnarray}\label{dpq}
&i\left[\pi^{ij}(x),h_{kl}(y)\right]=\frac{1}{2}\left(\delta_{k}^{i}\delta_{l}^{j}+\delta_{l}^{i}\delta_{k}^{j}\right)\delta^{(3)}(x,y),&\\
&i\left[\pi^i(x),N_j(y)\right]=\delta^i_j\delta^{(3)}(x,y)\quad,\quad i\left[\pi(x),N(y)\right]=\delta^{(3)}(x,y),&
\end{eqnarray}
yields the Wheeler--DeWitt equation \cite{whe, dew}
\begin{equation}\label{wdw}
\left\{G_{ijkl}\dfrac{\delta^2}{\delta h_{ij}\delta h_{kl}}+h^{1/2}\left({^{(3)}R}-2\Lambda-6\varrho\right)\right\}\Psi[h_{ij},\phi]=0,
\end{equation}
where $G_{ijkl}$ is the Wheeler metric on superspace $S(\partial M)$ \cite{whe, sup}
\begin{equation}
G_{ijkl}=\dfrac{1}{2\sqrt{h}}\left(h_{ik}h_{jl}+h_{il}h_{jk}-h_{ij}h_{kl}\right),
\end{equation}
and first class constraints are conditions on $\Psi[h_{ij},\phi]$
\begin{equation}
  \pi\Psi[h_{ij},\phi]=0\quad, \quad \pi^i\Psi[h_{ij},\phi]=0\quad,\quad H^i\Psi[h_{ij},\phi]=0.
\end{equation}

In fact quantum general relativity, given by the Wheeler--DeWitt equation, historically is one of the first attempts of quantum gravity theory construction. Actually, however, quantum geometrodynamics has became the motivation for development of the quantum gravity idea and building novel formulations. We are going to build the our model basing on the Wheeler--DeWitt theory (\ref{wdw}). Strictly speaking, however, the proposing toy model will possess a reductionist character.

\section{The Ansatz}\label{sec:2}
For construction of a simple quantum gravity model let us apply the following three step ansatz.

\indent\textbf{Global one--dimensionality conjecture.} Suppose that Matter fields are one-variable functionals $\phi=\phi[h]$ where $h$ is a volume form of $\partial M$
\begin{equation}
h\equiv\det h_{ij}=\dfrac{1}{3}\epsilon^{ijk}\epsilon^{lmn}h_{il}h_{jm}h_{kn},
\end{equation}
and $\epsilon$ is the Levi-Civita symbol. Also we assume, as an element of the model, that gravitational field is described only variable h. In result a wave function $\Psi[h_{ij},\phi]$ becomes
\begin{equation}\label{GOD}
  \Psi[h_{ij},\phi]\rightarrow\Psi[h],
\end{equation}
so that, the proposed quantum gravity model is
\begin{equation}\label{wdw1}
\left\{-G_{ijkl}\dfrac{\delta^2}{\delta h_{ij}\delta h_{kl}}-h^{1/2}\left({^{(3)}R}-2\Lambda-6\varrho[h]\right)\right\}\Psi[h]=0.
\end{equation}
In analogy to the generic cosmology \cite{gli}, (\ref{GOD}) describes isotropic spacetimes\footnote{Assumption (\ref{GOD}) means that we consider a strata of full superspace, \emph{i.e.} the DeWitt minisuperspace model where the wave function depends only one variable $h$.}.
\indent\textbf{Reduced quantum geometrodynamics.} Using $3+1$ splitting (\ref{dec}) within the Jacobi formula \cite{gr}
\begin{equation}\label{dg}
  \delta g = gg^{\mu\nu}\delta g_{\mu\nu},
\end{equation}
establishes the Jacobian matrix for for the reduction of variables $h_{ij}$ to $h$
 \begin{eqnarray}\label{hij}
  N^2\delta h=N^2hh^{ij}\delta h_{ij} \longrightarrow \mathcal{J}\left(h_{ij},h\right)=\dfrac{\delta (h)}{\delta (h_{ij})}=\dfrac{\delta h}{\delta h_{ij}}\equiv hh^{ij}.
\end{eqnarray}
Because of the approximation (\ref{GOD}) the variational derivative $\delta/\delta h_{ij}$ acts on functional depending only on $h$.
It allows us to express the derivative with respect  $h_{ij}$ through the derivative with respect $h$. Therefore
\begin{equation}\label{hij1}
\dfrac{\delta \Psi[h]}{\delta h_{ij}} = h h^{ij}\dfrac{\delta \Psi[h]}{\delta h}.
\end{equation}
Consequently, application of (\ref{hij1}) within the differential operator in (\ref{wdw1}) gives
\begin{eqnarray}\label{gij1}
 G_{ijkl}\dfrac{\delta^2}{\delta h_{ij}\delta h_{kl}}=G_{ijkl}h^{ij}h^{kl}h^2\dfrac{\delta^2}{\delta h^2}.
\end{eqnarray}
So that the reduction is given by the double contraction
\begin{eqnarray}\label{gij2}
  G_{ijkl}h^{ij}h^{kl}=\dfrac{1}{2\sqrt{h}}\left(h_{ik}h_{jl}+h_{il}h_{jk}-h_{ij}h_{kl}\right)h^{ij}h^{kl}=-\dfrac{3}{2}h^{-1/2},
\end{eqnarray}
where we have used the relations for 3-dimensional embedding $h^{ab}h_{bc}=h^a_c$, $h^a_a=\tr h_{ab}=3$. Jointing (\ref{gij1}) and (\ref{gij2}) one obtains finally the relation\footnote{Because the relation (\ref{hij}) arises due to $3+1$ approximation, so (\ref{gij2}) is an approximation within the ansatz.}
\begin{eqnarray}
 G_{ijkl}\dfrac{\delta^2}{\delta h_{ij}\delta h_{kl}}=-\dfrac{3}{2}h^{3/2}\dfrac{\delta^2}{\delta h^2},
\end{eqnarray}
which leads to the reduced theory
\begin{equation}\label{new}
  \left\{\dfrac{3}{2}h^{3/2}\dfrac{\delta^2}{\delta h^2}+h^{1/2}\left({^{(3)}R}-2\Lambda-6\varrho[h]\right)\right\}\Psi[h]=0.
\end{equation}
\indent \textbf{Dimensional reduction.} The model (\ref{new}) can be rewritten as
\begin{equation}\label{kgf}
\left(\dfrac{\delta^2}{\delta{h^2}}-m^2\right)\Psi=0,
\end{equation}
where $m^2$ is a squared (variable) mass of $\Psi$
\begin{equation}\label{masqr}
m^2=\dfrac{2}{3h}\left(^{(3)}R-2\Lambda-6\varrho\right)=\dfrac{2}{3h}(K_{ij}K^{ij}-K^2),
\end{equation}
and scalar constraint was used. Eq. (\ref{kgf}) arises by stationarity of the action\footnote{Here $S[\Psi]$ is a field-theoretic action functional in $\Psi$ so that any dependence on $h$ of the mass $m$ does not play a role for equations of motion $\delta S/\delta\Psi=0$.}
\begin{equation}\label{act1}
  S[\Psi]=\int \delta h L\left(\Psi,\dfrac{\delta \Psi}{\delta h}\right)\quad,\quad L=\dfrac{1}{2}\left(\Pi_\Psi^2+m^2\Psi^2\right)\quad,\quad \Pi_\Psi=\dfrac{\delta \Psi}{\delta h}.
\end{equation}
By using $\Pi_\Psi$ one rewrites the equation (\ref{kgf}) as
\begin{equation}\label{1}
  \dfrac{\delta\Pi_\Psi}{\delta h}-m^2\Psi=0,
\end{equation}
which together with $\Pi_\Psi$ in (\ref{act1}) yields the appropriate Dirac equation
\begin{equation}\label{dira}
 \left(i\gamma\dfrac{\delta}{\delta h}-M\right)\Phi=0\quad,\quad\Phi=\left[\begin{array}{c}\Pi_\Psi\\ \Psi\end{array}\right]\quad,\quad M=\left[\begin{array}{cc}
-1&0\\0&m^{2}\end{array}\right].
\end{equation}
The $\gamma$ matrices algebra consists only one element - the Pauli matrix $\sigma_y$
\begin{equation}\label{cliff}
  \gamma=\left[\begin{array}{cc}0&-i\\i&0\end{array}\right]\quad,\quad\gamma^2=I\quad,\quad\left\{\gamma,\gamma\right\}=2\delta_E\quad,\quad\delta_E=\left[\begin{array}{cc}1&0\\0&1\end{array}\right].
\end{equation}
The matrix algebra (\ref{cliff}) forms the Euclidean Clifford algebra \cite{euc} $\mathcal{C}\ell_{1,1}(\mathbb{R})$ which has a $2D$ complex representation. Restricting to $Pin_{1,1}(\mathbb{R})$ yield a complex representation of 2D Pin group (2D spin representation); restricting to $Spin_{1,1}(\mathbb{R})$ splits it onto a sum of two half spin $1D$ representations (1D Weyl representation). The algebra decomposes into a direct sum of central simple algebras isomorphic to matrix algebra over $\mathbb{R}$
\begin{eqnarray}
\mathcal{C}\ell_{1,1}(\mathbb{R})=\mathcal{C}\ell^+_{1,1}(\mathbb{R})\oplus \mathcal{C}\ell^-_{1,1}(\mathbb{R})\quad,\quad\mathcal{C}\ell_{1,1}(\mathbb{R})\cong\mathbb{R}(2)\quad,\quad\mathcal{C}\ell^{\pm}_{1,1}(\mathbb{R})\cong\mathbb{R},
\end{eqnarray}
and moreover has a decomposition into a tensor product
\begin{eqnarray}
  \mathcal{C}\ell_{1,1}(\mathbb{R})=\mathcal{C}\ell_{2,0}(\mathbb{R})\otimes\mathcal{C}\ell_{0,0}(\mathbb{R})\quad,\quad\mathcal{C}\ell_{0,0}(\mathbb{R})\cong\mathbb{R}.
\end{eqnarray}

\section{Quantization}\label{sec:3}
The Dirac equation (\ref{dira}) can be canonically quantized (See \emph{e.g.} \cite{cq})
\begin{eqnarray}
i\left[\mathbf{\Pi}_{\Psi}[h'],\mathbf{\Psi}[h]\right]=\delta(h'-h)~,~i\left[\mathbf{\Pi}_{\Psi}[h'],\mathbf{\Pi}_{\Psi}[h]\right]=0~,~i\left[\mathbf{\Psi}[h'],\mathbf{\Psi}[h]\right]=0.\label{c3}
\end{eqnarray}
Using of the Fock space allows to derive the solution in the form
\begin{equation}\label{sqx}
  \mathbf{\Phi}=\mathbb{Q}\mathfrak{B}\quad,\quad \mathbb{Q}=\dfrac{1}{\sqrt{2}}\left[\begin{array}{cc}\sqrt{1/|m|}&\sqrt{1/|m|}\\
-i\sqrt{|m|}&i\sqrt{|m|}\end{array}\right],
\end{equation}
where $\mathfrak{B}=\mathfrak{B}[h]$ is a dynamical reper
\begin{equation}\label{db}
  \mathfrak{B}=\left\{\left[\begin{array}{c}\mathsf{G}[h]\\
\mathsf{G}^{\dagger}[h]\end{array}\right]:\left[\mathsf{G}[h'],\mathsf{G}^{\dagger}[h]\right]=\delta\left(h'-h\right), \left[\mathsf{G}[h'],\mathsf{G}[h]\right]=0\right\},
\end{equation}
and yields non Heisenberg-like dynamics of (\ref{db})
\begin{equation}\label{df}
\dfrac{\delta\mathfrak{B}}{\delta h}=\mathbb{X}\mathfrak{B}\quad,\quad\mathbb{X}=\left[\begin{array}{cc}
-im&\dfrac{1}{2m}\dfrac{\delta m}{\delta h}\\
\dfrac{1}{2m}\dfrac{\delta m}{\delta h}&im\end{array}\right].
\end{equation}
Supposing that there is an other reper $\mathfrak{F}$ determined by the Bogoliubov transformation and the Heisenberg equations of motion
\begin{eqnarray}
\mathfrak{F}&=&\left[\begin{array}{cc}u&v\\v^{\ast}&u^{\ast}\end{array}\right]\mathfrak{B}, \qquad |u|^2-|v|^2=1,\label{pr2}\\
\dfrac{\delta\mathfrak{F}}{\delta h}&=&\left[\begin{array}{cc}-i\Omega&0\\0&i\Omega\end{array}\right]\mathfrak{F},\label{pr3}
\end{eqnarray}
where $u$, $v$, $\Omega$ are functionals of $h$, one obtains
\begin{equation}\label{bcof}
  \dfrac{\delta\mathbf{b}}{\delta h}=\mathbb{X}\mathbf{b}\quad,\quad\mathbf{b}=\left[\begin{array}{c}u\\v\end{array}\right],
\end{equation}
and $\Omega\equiv0$, so that $\mathfrak{F}$ is the Fock static reper with respect to initial data ($I$)
\begin{equation}\label{in}
\mathfrak{F}=\left\{\left[\begin{array}{c}\mathsf{G}_I\\
\mathsf{G}^{\dagger}_I\end{array}\right]: \left[\mathsf{G}_I,\mathsf{G}^{\dagger}_I\right]=1, \left[\mathsf{G}_I,\mathsf{G}_I\right]=0\right\},
 \end{equation}
and vacuum state $\vac$ is correctly defined
\begin{equation}
\mathsf{G}_I\vac=0\quad,\quad 0=\ivac \mathsf{G}_I^\dagger.
\end{equation}
Integrability of Eqs. (\ref{bcof}) is crucial. The transformation (\ref{pr2}) suggests employing the superfluid parametrization which yield\footnote{In (\ref{sup}) the functional measure $\delta h$ for the case of a fixed space configuration transits into the Riemann--Lebesgue measure $dh$. However, $h$ in general is a smooth function of space parameters, $\delta h$ is a total variation and has a sense of the Stieltjes measure.}
\begin{equation}
u=\dfrac{1+\lambda}{2\sqrt{\lambda}}\exp\left\{im_I\int_{h_I}^{h}\dfrac{\delta h'}{\lambda'}\right\}\quad,\quad v=\dfrac{1-\lambda}{2\sqrt{\lambda}}\exp\left\{-im_I\int_{h_I}^{h}\dfrac{\delta h'}{\lambda'}\right\},\label{sup}
\end{equation}
where $\lambda\equiv\lambda[h]$, $\lambda'=\lambda[h']$ is a length scale \emph{i.e.} inverted mass scale $\mu=m/m_I=1/\lambda$. Consequently the integrability problem is solved by \begin{equation}\label{phi}
\mathbf{\Phi}=\mathbb{Q}\mathbb{G}\mathfrak{F},
\end{equation}
where $\mathbb{G}$ is the monodromy matrix
\begin{equation}\label{mono}
\mathbb{G}=\left[\begin{array}{cc}
\dfrac{\lambda+1}{2\sqrt{\lambda}}\exp\left\{-im_I\int_{h_I}^{h}\dfrac{\delta h'}{\lambda'}\right\}\vspace*{10pt}&
\dfrac{\lambda-1}{2\sqrt{\lambda}}\exp\left\{im_I\int_{h_I}^{h}\dfrac{\delta h'}{\lambda'}\right\}\\
\dfrac{\lambda-1}{2\sqrt{\lambda}}\exp\left\{-im_I\int_{h_I}^{h}\dfrac{\delta h'}{\lambda'}\right\}&
\dfrac{\lambda+1}{2\sqrt{\lambda}}\exp\left\{im_I\int_{h_I}^{h}\dfrac{\delta h'}{\lambda'}\right\}\end{array}\right].
\end{equation}

One sees now that the presented model expresses quantum gravity as a quantum field theory, where the quantum field associated with a space configuration is given by the relation (\ref{phi}). In this manner one can write out some straightforward conclusions following form the simple model.
\section{Some implications of general formulation}\label{sec:4}
The proposed field-theoretic model was solved. However, still we do not know what it the role of the 1D wave function given by the equation (\ref{wdw1}). The same problem is to define any geometric quantities related to this model. The quantum field theory (\ref{phi}) has also unclear significance. Let us present now some conclusions arising from the previous section's model, which will clarify these questions in some detail.\\
\indent\textbf{Global 1D wave function.} The Dirac equation (\ref{dira}) can be rewritten in the form of Schr\"odinger-like evolution equation
\begin{equation}\label{evol}
  \dfrac{\delta\Phi}{\delta h}=H\Phi\quad,\quad H=-\left[\begin{array}{cc}0&\dfrac{m_I^2}{\lambda^2}\\1&0\end{array}\right],
\end{equation}
yielding unitary evolution operator $U=U(h,h_I)=\exp \int_{h_I}^h H\delta h$ given by
\begin{equation}
U=\left[\begin{array}{cc}\cosh f[h,h_I]&\left(-m_I^2\int_{h_I}^h\dfrac{\delta h'}{\lambda'^2}\right)\dfrac{\sinh f[h,h_I]}{f[h,h_I]}\\
(h_I-h)\dfrac{\sinh f[h,h_I]}{f[h,h_I]}&\cosh f[h,h_I]\end{array}\right],
\end{equation}
where $f[h,h_I]=|m_I|\sqrt{\strut{(h-h_I)\int_{h_I}^h\frac{\delta h'}{\lambda'^2}}}$, so that Eq. (\ref{evol}) is solved by
\begin{equation}
  \Phi[h]=U(h,h_I)\Phi[h_I].
\end{equation}
Straightforward elementary algebraic manipulations allow to determine the global one-dimensional wave function as
\begin{equation}\label{wfun}
  \Psi=\Psi^I\cosh f[h,h_I]-\Pi_\Psi^I(h-h_I)\sgn(h-h_I)\dfrac{\sinh f[h,h_I]}{f[h,h_I]},
\end{equation}
and similarly the canonical conjugate momentum is
\begin{equation}\label{pfun}
  \Pi_\Psi=\Pi_\Psi^I\cosh f[h,h_I]-\Psi^Im_I^2\sgn(h-h_I)\left(\int_{h_I}^h\dfrac{\delta h'}{\lambda'^2}\right)\dfrac{\sinh f[h,h_I]}{f[h,h_I]},
\end{equation}
where $\Psi^I=\Psi[h_I]$ and $\Pi_\Psi^I=\Pi_{\Psi}[h_I]=\left.\dfrac{\delta\Psi}{\delta h}\right|_{h=hI}$ are initial data. In this manner the probability density in the classical reduced model is
\begin{eqnarray}
  \Omega[h]&=&(\Psi^I)^2\cosh^2 f[h,h_I]+(\Pi_\Psi^I)^2(h-h_I)^2\left[\dfrac{\sinh{f[h,h_I]}}{f[h,h_I]}\right]^2-\nonumber\\
  &-&2\Psi^I\Pi_\Psi^I(h-h_I)\sgn(h-h_I)\dfrac{\sinh{2f[h,h_I]}}{2f[h,h_I]},
\end{eqnarray}
and $\Psi^I$ and $\Pi_\Psi^I$ are determined by the normalization condition
\begin{equation}\label{abc}
  \int_{h_I}^\infty\Omega[h']\delta h'=1\quad\longrightarrow\quad C(\Pi_\Psi^I)^2-2B\Psi^I\Pi_\Psi^I+A(\Psi^I)^2-1=0,
\end{equation}
where the constants $A$, $B$, $C$ are given by the integrals
\begin{eqnarray}
  A&=&\int_{h_I}^\infty\cosh^2 f[h',h_I]\delta h',\\
  B&=&\int_{h_I}^\infty(h'-h_I)\sgn(h'-h_I)\dfrac{\sinh{2f[h',h_I]}}{2f[h',h_I]}\delta h',\\
  C&=&\int_{h_I}^\infty(h'-h_I)^2\left[\dfrac{\sinh{f[h',h_I]}}{f[h',h_I]}\right]^2\delta h',
\end{eqnarray}
The equation (\ref{abc}) can be solved straightforwardly. In result one obtains
\begin{eqnarray}\label{pii}
\Pi_\Psi^I=\dfrac{B}{C}\Psi^I\pm\sqrt{\strut{\left[\left(\dfrac{B}{C}\right)^2-\dfrac{A}{C}\right](\Psi^I)^2+\dfrac{1}{C}}}.
\end{eqnarray}
Using $\Pi_\Psi^I=\dfrac{\delta \Psi^I}{\delta h_I}$ in (\ref{pii}) yields differential equation for $\Psi^I$, with the solution
\begin{equation}\label{wf}
  \Psi^I=f_{\pm}^{(-1)}\left(\pm\dfrac{h_I}{C}+C_1\right),
\end{equation}
where $C_1$ is integration constant, and $f_{\pm}(h_I)$ are the functions
\begin{eqnarray}\label{func}
  \!\!\!\!\!\!\!\!\!\!\!\!\!\!\!\!\!\!\!\!\!\!\!\!\!\!&&f_{\pm}(h_I)=\dfrac{B}{AC}\Bigg\{ \artanh\dfrac{Bh_I}{\sqrt{\strut{\left(B^2-AC\right)h_I^2+C}}}\pm\ln\sqrt{\strut{Ah_I^2-1}}-\nonumber\\
  \!\!\!\!\!\!\!\!\!\!\!\!\!\!\!\!\!\!\!\!\!\!\!\!\!\!&&-\dfrac{\sqrt{\strut{B^2-AC}}}{B}\ln\left[\left(B^2-AC\right)h_I+\sqrt{\strut{B^2-AC}}\sqrt{\strut{\left(B^2-AC\right)h_I^2+C}}\right]\Bigg\}.
\end{eqnarray}

\indent\textbf{The 3-dimensional manifolds.} The model (\ref{kgf}) can be rewritten as
\begin{equation}\label{kgfq1}
\left(\dfrac{\delta^2}{\delta{h^2}}-\dfrac{2}{3h}{^{(3)}R}\right)\Psi[h]=-\dfrac{4}{h}\left(\varrho[h]+\dfrac{\Lambda}{3}\right)\Psi[h].
\end{equation}
and considered as the equation for the 3-dimensional scalar curvature ${^{(3)}\!R}$
\begin{equation}\label{kgfq2}
^{(3)}\!R=-6\left(\varrho[h]+\dfrac{\Lambda}{3}\right)+\varphi(\Psi)h\quad,\quad  \varphi(\Psi)=\dfrac{3}{2}\dfrac{1}{\Psi}\dfrac{\delta^2\Psi}{\delta{h^2}}.
\end{equation}
In the vacuum case, \emph{i.e.} for the conditions $\left(\varrho[h]\equiv0 \cap \Lambda\equiv0\right)$ or $\varrho[h]=-\dfrac{\Lambda}{3}$,
one obtains from (\ref{kgfq2}) that
\begin{equation}\label{kgfq3}
^{(3)}\!R=\varphi_nh,
\end{equation}
where $\varphi_n$ in an eigenvalue determined by the equation
\begin{equation}
  \dfrac{\delta^2\Psi}{\delta{h^2}}-\dfrac{2}{3}\varphi_n\Psi=0.
\end{equation}
Supposing analytical form of $\Psi$ one establishes $\varphi_n$
\begin{equation}
  \Psi=\sum_{n=0}^{\infty}a_n(h-h_I)^n\longrightarrow\varphi_n=\dfrac{3}{2}\left.\left(\dfrac{\delta ^{n}}{\delta h^{n}}\left(\dfrac{m_I^2}{\lambda^2[h]}\Psi[h]\right)\left/\dfrac{\delta ^{n}\Psi[h]}{\delta h^{n}}\right.\right)\right|_{h=h_I}.\label{vph}
\end{equation}
Let us assume that there are generalized functional Fourier transforms
\begin{equation}
  \widetilde{\Psi}[s]=\int\delta h e^{-2\pi ish}\Psi[h]\quad,\quad\widetilde{\dfrac{1}{\lambda^2}}[s]=\int\delta h e^{-2\pi ish}\dfrac{1}{\lambda^2[h]},\label{ft2}
\end{equation}
as well as the generalized Leibniz product formula for functional derivatives
\begin{equation}
  \dfrac{\delta ^{n}}{\delta h^{n}}\left(\dfrac{1}{\lambda^2[h]}\Psi[h]\right)=\sum_{r=0}^n\binom{n}{r}\left(\dfrac{\delta^r}{\delta h^r}\dfrac{1}{\lambda^2[h]}\right)\left(\dfrac{\delta^{n-r}}{\delta h^{n-r}}\Psi[h]\right).\label{leib}
\end{equation}
Using (\ref{ft2}), (\ref{leib}) and $\sum_{r=0}^n\binom{n}{r}x^r=(1+x)^n$, within (\ref{vph}) yields
\begin{equation}\label{vp1}
  \varphi_n=\dfrac{3}{2}m_I^2\iint\delta s'\delta se^{2i\pi(s'+s)h_I}\left(1+\dfrac{s'}{s}\right)^n\widetilde{\dfrac{1}{\lambda^2}}[s']\widetilde{\Psi}[s],
\end{equation}
so that applying the inverted Fourier transforms
\begin{equation}
  \Psi[h]=\int\delta s e^{2\pi ish}\widetilde{\Psi}[s]\quad,\quad\dfrac{1}{\lambda^2[h]}=\int\delta s e^{2\pi ish}\widetilde{\dfrac{1}{\lambda^2}}[s],\label{ft2a}
\end{equation}
within the relation (\ref{vp1}) one receives finally
\begin{eqnarray}\label{vp2}
\!\!\!\!\!\!\!\!\!\!\varphi_n&=&\dfrac{3}{2}\iint\delta h\delta h\mathcal{G}(h-h_I)\dfrac{m_I^2}{\lambda^2[h]}\Psi[h]=\dfrac{3}{2}\iint\delta h\delta h\mathcal{G}(h-h_I)\dfrac{\delta^2\Psi[h]}{\delta h^2}=\nonumber\\
\!\!\!\!\!\!\!\!\!\!&=&\dfrac{3}{2}\iint\delta h\delta h\mathcal{G}(h-h_I)\dfrac{\delta\Pi_\Psi[h]}{\delta h}=-\dfrac{3}{2}\iint\delta h\delta h\dfrac{\delta}{\delta h}\mathcal{G}(h-h_I)\Pi_\Psi[h],
\end{eqnarray}
where we have used equations of motion, partial integration method. In (\ref{vp2}) the kernel $\mathcal{G}(h-h_I)$ and its derivative can be derived straightforwardly as
\begin{eqnarray}
  \mathcal{G}(h-h_I)&=&\iint\delta s'\delta se^{-2i\pi(s'+s)(h-h_I)}\left(1+\dfrac{s'}{s}\right)^n,\label{kern1}\\
  \dfrac{\delta}{\delta h}\mathcal{G}(h-h_I)&=&-\iint\delta s'\delta s\dfrac{e^{-2i\pi(s'+s)(h-h_I)}}{2i\pi(s'+s)}\left(1+\dfrac{s'}{s}\right)^n.\label{kern2}
\end{eqnarray}
Estimation of the functional integrals (\ref{kern1}) or (\ref{kern2}), and using of (\ref{wfun}) or (\ref{pfun}), leads to (\ref{vp2}), which is a crucial for the relation (\ref{kgfq3}).

\indent\textbf{Quantum correlations.} With using of the matrices (\ref{mono}) and (\ref{sqx}), and the relation (\ref{phi}) one derives the quantum field
\begin{eqnarray}\label{field}
  \mathbf{\Psi}[h]=\frac{\lambda[h]}{2\sqrt{2m_I}}\left(\exp\left\{-im_I\int_{h_I}^h\dfrac{\delta h'}{\lambda[h']}\right\}\mathsf{G}_I+\exp\left\{im_I\int_{h_I}^h\dfrac{\delta h'}{\lambda[h']}\right\}\mathsf{G}_I^\dagger\right).
\end{eqnarray}
Taking into account the $n$-field one-point quantum states determined as
\begin{eqnarray}
|h,n\rangle\equiv\mathbf{\Psi}^n\vac=\left(\frac{\lambda}{2\sqrt{2m_I}}\exp\left\{im_I\int_{h_I}^h\dfrac{\delta h'}{\lambda'}\right\}\right)^n\mathsf{G}^{\dagger n}_I\vac,
\end{eqnarray}
yields two-point correlators $\cor_{n'n}(h',h)\equiv\langle n',h'|h,n\rangle$ or explicitly
\begin{eqnarray}\label{gencor}
  \cor_{n'n}(h',h)&=&\dfrac{\lambda^{\prime n'}\lambda^n}{\left(\sqrt{8m_I}\right)^{n'+n}}\exp\left\{im_I\left(n'\int_{h'}^{h_I}+n\int_{h_I}^{h}\right)\dfrac{\delta h''}{\lambda^{\prime\prime}}\right\}\times\nonumber\\
  &\times&\ivac\mathsf{G}_I^{n'}\mathsf{G}^{\dagger n}_I\vac,\label{cor0}
\end{eqnarray}
where $\lambda^\prime\equiv\lambda[h']$ and so on. Basically one obtains
\begin{eqnarray}
\!\!\!\!\!\!\!\!\!\!\!\!\!\!\!\!\!\!&&\cor_{00}(h,h)=\cor_{00}(h',h)=\cor_{00}(h_I,h_I)=\left\langle \mathrm{VAC}|\mathrm{VAC}\right\rangle,\\
\!\!\!\!\!\!\!\!\!\!\!\!\!\!\!\!\!\!&&\cor_{11}(h_I,h_I)=\dfrac{1}{8m_I}\quad,\quad \dfrac{\cor_{n'n}(h_I,h_I)}{\left[\cor_{11}(h_I,h_I)\right]^{(n'+n)/2}}=\ivac\mathsf{G}_I^{n'}\mathsf{G}^{\dagger n}_I\vac,
\end{eqnarray}
so by elementary algebraic manipulations one receives
\begin{eqnarray}
\!\!\!\!\!\!\!\!\!\!\!\!\!\!\!\!\!\!&&\cor_{11}(h',h)=\dfrac{\sqrt{\strut{\cor_{11}(h',h')\cor_{11}(h,h)}}}{\cor_{11}(h_I,h_I)}\exp\left\{im_I\int_{h'}^{h}\dfrac{\delta h''}{\lambda''}\right\},\\
\!\!\!\!\!\!\!\!\!\!\!\!\!\!\!\!\!\!&&\dfrac{\cor_{nn}(h',h)}{\cor_{00}(h_I,h_I)}=\left[\dfrac{\cor_{11}(h',h)}{\cor_{00}(h_I,h_I)}\right]^n\quad,\quad\dfrac{\cor_{11}(h,h)}{\cor_{00}(h_I,h_I)}=\lambda^2\cor_{11}(h_I,h_I).\label{cor2}
\end{eqnarray}
Straightforwardly from (\ref{cor2}) one relate a size scale with quantum correlations
\begin{eqnarray}
\lambda=\sqrt{\strut{\dfrac{\cor_{11}(h,h)}{\cor_{11}(h_I,h_I)\cor_{00}(h_I,h_I)}}},\label{lam}
\end{eqnarray}
that consequently leads to the formulas
\begin{equation}
\dfrac{\cor_{n'n}(h,h)}{\cor_{n'n}(h_I,h_I)}=\lambda^{n'+n}\exp\left\{-im_I(n'-n)\int_{h_I}^h\dfrac{\delta h'}{\lambda'}\right\},\label{cor4}
\end{equation}
\begin{equation}
\dfrac{\cor_{11}(h',h)}{\cor_{00}(h_I,h_I)\cor_{11}(h_I,h_I)}=\lambda'\lambda\exp\left\{im_I\int_{h'}^{h}\dfrac{\delta h''}{\lambda''}\right\},
\end{equation}
\begin{equation}
\dfrac{\cor_{nn}(h',h)}{\cor_{00}(h_I,h_I)}=\lambda'^n\lambda^n[\cor_{11}(h_I,h_I)]^n\exp\left\{im_In\int_{h'}^{h}\dfrac{\delta h''}{\lambda''}\right\}.
\end{equation}
A whole information on the system is contained in $\lambda$, $\mu$, and $m_I$. Quantum correlations are determined by these quantities only. By measurement of quantum correlations one deduces $\lambda$, $\mu$, $m_I$.

The presented conclusions have a formal character, however, they show a general feature of the proposed simple model of quantum gravity. We have solved the model by the 1D wave function (\ref{wfun}). We have discussed the 3-dimensional manifolds (\ref{kgfq3}) corresponding to this model, and we have found the relation between quantum correlations and physical scales (\ref{lam}).

\section{Summary}\label{sec:5}

This paper discussed some consequences of the global one--dimensionality conjecture within the Wheeler--DeWitt theory. We have applied a field theory for formulation of the simple model of quantum gravity. The model was constructed by the following steps
\begin{enumerate}
    \item We have started from general relativity of compact manifold with boundaries; its action was written in $3+1$ splitting and the scalar constraint was canonically quantized. Resulting theory was the Wheeler--DeWitt model of quantum gravity;
    \item Next stage we have proposed the ansatz based on the global one-dimensionality conjecture;
    \item With using of the ansatz the quantum geometrodynamics was reduced to one-dimensional global evolution, with the dimension being an embedding volume form;
    \item Employing canonical formalism, we have used a field-theoretic action corresponding to the model, and by dimensional reduction the appropriate Dirac equation was received;
    \item Finally the Fock quantization was applied. Static reper of creators and annihilators was found by using of the diagonalization procedure employing the Bogoliubov transformation and the Heisenberg equations of motion. Consequently, the proposed model is defining quantum gravity as a quantum field theory. The quantum field was derived in a straightforward way (\ref{phi}).
\end{enumerate}
In result, we have obtained correctly defined integrability problem, which allowed to study its formal consequences. Particularly, we have discussed
\begin{enumerate}
  \item The integrability problem and global one-dimensional wave function. It was shown that by integration of the model in the Schr\"odinger equation form there is a possibility to obtain an exact solutions of the model.
  \item 3-dimensional manifolds corresponding to the model. It was shown that the model is defining a 3-dimensional manifolds $^{(3)}\!R=\phi_nh$, where for given wave function the parameter $\phi_n$ can be derived by the assumption of the appropriate Fourier transforms and its inverted transforms.
  \item Relation between quantum correlations and physical scales. We have connected one-point quantum correlations with size and mass scales.
  \end{enumerate}
The presented conclusions are partial, but they show possible physical and geometric implications following from the simple quantum gravity model. From a mathematical point of view we have applied a strategy of one-dimensional integration, with the Riemann--Lebesgue measure for fixed space configuration or the Stieltjes measure in general case. Both physical and mathematical sides of the model were emphasized in this paper.

\section*{Acknowledgements}
Special thanks are directed to Prof. {I. L. Buchbinder} for very constructive discussion and his helpful comments to primary notes of the author. The author benefitted also many valuable discussions from Profs. A. B. Arbuzov, {I. Ya. Aref'eva}, B. M. Barbashov, {K. A. Bronnikov}, and V. N. Pervushin.

\end{document}